\documentstyle[pra,aps]{revtex}
\input epsf
\begin{document}
\def\question#1{{{\marginpar{\tiny \sc #1}}}}
\draft
\title{Can Geodesics in Extra Dimensions Solve the Cosmological
Horizon Problem?}
\author{Daniel J. H. 
Chung$^{1,}$\thanks{Electronic mail: {\tt
djchung@umich.edu}} and Katherine Freese$^{1,2,}$\thanks{Electronic mail: {\tt
ktfreese@umich.edu}}} 
\address{$^1$ Randall Physics Laboratory,
	 University of Michigan, Ann Arbor, MI 48109-1120} 
\address{$^2$ Max Planck Institut fuer Physik,
	Foehringer Ring 6, 80805 Muenchen, Germany} 
\date{September 1999}
 \maketitle

 \begin{abstract}
 We demonstrate a non-inflationary 
 solution to the cosmological horizon 
 problem in scenarios in which our observable universe is
 confined to three spatial dimensions (a three-brane) embedded in a
 higher dimensional space.  A signal traveling along an extra-dimensional null
 geodesic may leave our three-brane, travel into the extra dimensions,
 and subsequently return to a different place on our three-brane in a
 shorter time than the time a signal confined to our three-brane would take.
 Hence, these geodesics may connect distant points which would
 otherwise be ``outside'' the four dimensional horizon (points not in
 causal contact with one another).
 \end{abstract}

 \def\question#1{{{\marginpar{\tiny \sc #1}}}}
 \def\eqr#1{{Eq.\ (\ref{#1})}}
 \def\be{\begin{equation}}
 \def\ee{\end{equation}}

 \section{Introduction}
 The universe appears to be homogeneous and isotropic on large scales.
 According to the COBE measurements, the cosmic background radiation
(CBR) is
 uniform to a part in $10^4$ on large scales (from about $10''$ to $180
 \deg$ \cite{kolbturner}).  Furthermore, the light element abundance
 measurements seem to indicate that the observable universe (bounded by
 the last scattering surface) was homogeneous by the time of
 nucleosynthesis \cite{yangetal}.  Hence, we would expect the
 observable universe today (time $t_0$) to have been in causal contact
 by the time of nucleosynthesis $t_n$; otherwise the initial conditions
 of the universe would have to be extremely fine tuned in order for
 the causally disconnected patches to resemble one another as much as
 they do.  However, in a Friedmann Robertson Walker (FRW) universe (a
metric of $ds^2= dt^2 - a(t)^2 d\bf{x}^2$)
 that is matter or radiation dominated, upon naive extrapolation back
 to the singularity, one finds that there is a finite horizon length at
 the time of nucleosynthesis.  Hence, for the observable universe to
 have been in causal contact by the time of nucleosynthesis, the
 comoving horizon length must have been larger than the comoving
 distance to the last scattering surface.  In other words,
 our observable universe today (when appropriately scaled back
 to the time of nucleosynthesis) must have fit inside a causal region
 at the time of nucleosynthesis.

 The comoving size $L_o$ of the observable universe today is
 \begin{equation} \label{sizeobs} L_o = \int_{t_{dec}}^{t_0}
 \frac{dt}{a(t)} \end{equation} where $t_{dec}$ is the time of the
 radiation decoupling and $t_0$ is the time today (subscript $0$
 refers to today).  The comoving size $L_n$ of the horizon at the time
 of nucleosynthesis is \begin{equation} \label{eq:horizonlength} L_n =
 \int_0^{t_n} \frac{dt}{a(t)}.  \end{equation} In order to explain
 causal contact of all points within our observable universe at the
 time of nucleosynthesis, we require $L_o < L_n$.  However, this
 condition is not met in a naive FRW cosmology with matter or
 radiation domination.  Even if we take $t_n$ to be the the time of
 last scattering of CBR and not the nucleosynthesis time, we still
 have a horizon problem by a factor of $10^5$.  In both matter or
 radiation domination cases, the time dependence of the scale factor
 is a power law with the index less than 1; in a dust (matter) dominated
 universe, $a \propto t^{2/3}$ and in a radiation dominated universe,
 $a \propto t^{1/2}$.  Hence, in the naive FRW cosmology, $L_0 \sim
 t_0/a_0$ and $L_n \sim t_n/a(t_n)$, such that $L_o > L_n$ while
 causal connection requires $L_o < L_n$.  This is the horizon problem.
 Inflationary cosmology \cite{guth} solves the horizon problem
 by having a period of accelerated expansion, with $\ddot a > 0$ (a
period of time when the universe was not dust or radiation dominated).

Here we consider a non-inflationary solution to the horizon problem.
The argument in the previous paragraph that leads to $L_o > L_n$ has
assumed that causal signals travel within the light-cone defined by
the null geodesics of a 4-dimensional manifold.  If the causal signals
can instead travel through higher dimensions, the points that are
seemingly causally disconnected from the 4-dimensional point of view
may in fact be causally connected.  A signal along the geodesic may
leave our spatially three dimensional world, travel into the extra
dimensions, and subsequently return to a different place in our three
dimensional world; the distance between the initial and the final
(return) point when measured along the 3-spatial dimensions may be
longer than the distance that a light signal confined to our 3+1
dimensional universe would travel in the same amount of time.  Such a
geodesic may arise when the curvature allows the path length for a
null signal (e.g., light ray) through the higher dimensions to be
shorter than any path length in our lower dimensional world alone.
Such a possibility has been alluded to before (see for example Ref.\
\cite{Kalbermann} and the footnote in Ref.\ \cite{rubakov}).  In this
paper, we construct explicit examples of such spacetimes.  Other
papers of interest relating to this topic can be found in Ref.\
\cite{moffat}, Refs.\ \cite{visser,cline}, and references therein.

In this paper, we will focus on noncompact 3-branes.  We construct an
example (in Section \ref{sec:patches}) which is compatible with the
cosmology of our universe.  In this scenario, there are two separate
3-branes: one is our observable universe and the other is the hidden
sector.  We take these 3-branes to be `parallel' to one another
(i.e. each of the branes is located at a constant coordinate value of
the extra dimension).  A field signal originating on our observable
3-brane travels away from our brane on a path perpendicular to both
branes and arrives at the other 3-brane.  There, it interacts with
fields confined to the hidden sector.  Subsequently, due to these
interactions, the signal travels along the hidden sector
3-brane. Because of the specific metric we have constructed, the
signal can traverse a longer coordinate distance than it could on our
brane in the same amount of time.  The signal then returns back to our
brane through a path perpendicular to the two 3-branes.  As a
consequence of this path, the signal has traversed an effective
distance on our 3-brane that is much longer than any distance it could
have covered had it remained on our 3-brane in the same time.  Hence
points outside of the naive 3-brane `horizon' can be connected in this
way.

 In Section \ref{sec:patches}, we describe a class of metrics which
 may be used to obtain semirealistic cosmology and for which the
 geodesic is higher dimensional: as described in the last paragraph,
 this model requires interactions between our brane and the hidden
 sector brane.  
In Section \ref{sec:embedd21}, we will
 construct a 2+1 dimensional example of a continuous metric (no
 interactions required) that allows geodesics to connect seemingly
 distant points; expansion of the universe is not yet taken into
 account in this simple case.  Then, in Section \ref{sec:onecurve}, we
 will generalize such a continuous metric to 4+1 dimensions with
 expansion.  However, the continuous case in Section
 \ref{sec:onecurve} does not produce a viable cosmology for our
 universe: first, the universe is not homogeneous (a special point is
 required and in this example is singular), and second, in this model
both our 3-dimensional world as 
 well as the bulk describing the extra dimensions are expanding with
 the same scale factor.  In Section \ref{sec:physical} we discuss some
 issues of causality violation.  Finally, in Section
 \ref{sec:conclude}, we summarize and conclude.

\newpage

 \section{Horizon Evading Metrics}
 \label{sec:metrics}

 \subsection{4+1 Dimensional Example with Viable Cosmology}
 \label{sec:patches}

 \begin{figure}
\hspace*{175pt} \epsfxsize=180pt \epsfbox{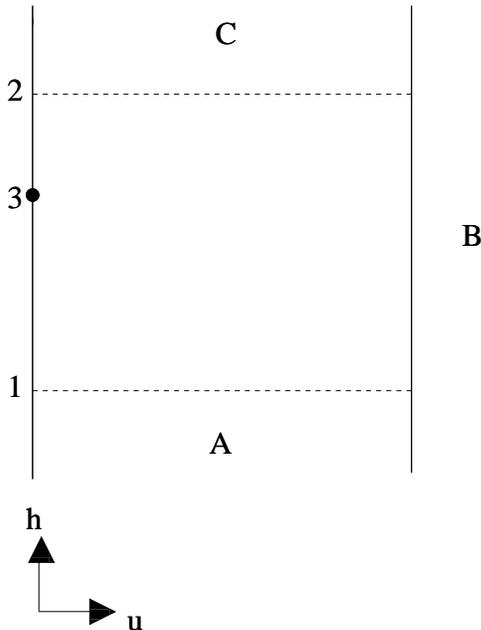}
 \caption{Branes and geodesics for 4+1 dimensional example.
 Our brane is represented by the left hand vertical line
 with $u=0$; a second brane is represented by the right hand
 vertical line with $u=L$.  The geodesic in the full metric
 leaves our brane at point 1, travels along A, B, and C, and
 reenters our brane at point 2.  The distance $h_{(1,3)}$ between
 points 1 and 3 is the horizon distance usually calculated in
 cosmology in the absence of extra dimensions.  Since
 $h_{(1,2)} > h_{(1,3)}$, points traditionally ``outside
 the horizon" are here causally connected.}
 \label{fig:best}
 \end{figure}

 Here we consider a 4+1 dimensional case which produces
 a viable cosmology: our observable universe is homogeneous
 and expanding.
 Consider a metric of the form
 \begin{equation}
 ds^2= dt^2- \left[ e^{-2 k u} a^2(t) d{\bf h}^2 + du^2 \right]
 \label{eq:simplesol}
 \end{equation}
 with our observable brane located at $u=0$.  Here
  ${\bf h}$ is a three dimensional Euclidean vector.  [Note that
 although this metric is similar to the one that was considered by
 \cite{randsund}, there is a crucial difference in that $dt^2$ term
 does not share the conformal factor multiplying $d{\bf h}^2$].  Now,
 consider the null geodesics labeled A, B, C, and D as shown in Fig.\
 \ref{fig:best}.  Explicitly, the null geodesics can be written as
 $(u=u_g(t), h=h_g(t))$ with
\begin{equation}
 u_g(t) =  
\left\{ 
\begin{array}{lll}
A: & t, & 0 \leq t \leq L \\
B: & L, & L \leq t \leq t_f-L \\
C: & t_f-t, & t_f-L \leq t \leq t_f
\end{array} 
\right.
\end{equation}
 and 
\begin{equation}
 h_g(t) =  
\left\{ 
\begin{array}{lll}
A: & 0, & 0 \leq t \leq L \\
B: & e^{k L} \int_L^{t-L} \frac{dt'}{a(t')} , & L \leq t \leq t_f-L \\ 
C: &  L , & t_f-L \leq t \leq t_f\\
\end{array} 
\right.
\end{equation}
with $\bf h$ chosen along a particular direction $h$ with an initial
value of $0$ without any loss of generality.  

One can show that, indeed, these paths labeled A,B, and C
satisfy the geodesic equations,
\begin{equation}
\frac{d^2 x^\mu}{d t^2} + \Gamma^\mu_{\alpha \beta} \frac{d x^\alpha}{d
t} \frac{d x^\beta}{d t} = 0
\label{eq:geodesic}
\end{equation}
where $\Gamma^{\mu}_{\alpha \beta}$ is the Christoffel symbol.  Here,
a signal originating on our observable 3-brane at $u=0$ travels away
from our brane on a path A perpendicular to the 3-brane.  Once it
arrives at $u=L$, it follows a trajectory B with constant
$u$. Subsequently it returns to our brane via trajectory C, again
perpendicular to our brane.  Hence the effective distance it traverses
on our brane is approximately given by the pathlength B (up to small
corrections).
  
The distance traveled by a null signal along the brane
between points 1 and 3 in time $t_f$ is 
\begin{equation}
h_{(1,3)} = \int_0^{t_f} {dt \over a} \, .
\end{equation}
As before, this is the distance that one would naively
interpret as the horizon size in eqn.(\ref{sizeobs}).\footnote{Note
that this does not correspond to a geodesic in the 5-dimensions.}
However, for a null signal leaving and reentering the brane via
the geodesic A, B, and C, the effective distance
traveled on the observable brane in time $t_f$ is
the distance between points 1 and 2, i.e., the same
as the distance traveled along path B: 
\begin{equation}
h_{(1,2)} = e^{k L} \int_L^{t_f-L} \frac{dt}{a(t)}.
\end{equation}
Clearly for large enough $kL$,
\begin{equation}
h_{(1,2)} > h_{(1,3)}
\end{equation} 
and regions that ordinarily would be considered out of causal contact
can be connected.
In the ordinary FRW cosmology, the surface of last scattering
of photons encompasses $10^5$ causally disconnected patches.
Here, however, if we take $t_f$ to be the time of last scattering 
of photons,
as long as $kL \sim \ln(10^5)$, then these patches
can have all been in contact with one another and 
we can solve the horizon problem.  
Because the induced metric on the brane in this scenario is
homogeneous, this scenario is cosmologically viable.

We have performed a numerical exploration of the solutions to the
geodesic equations in \eqr{eq:geodesic} resulting from the metric in
\eqr{eq:simplesol} for various initial conditions, in particular for
initial velocity vectors leaving the brane in a variety of directions.
We indeed find that there are geodesics with a continuous path which
leave and subsequently reenter our brane, i.e., there are
extra-dimensional causal paths which connect points 1 and 2 without a
need to jump from one geodesic to another (such as turning the corner
from segment A to segment B).  However we have not found continuous
paths which return to our brane at a point more distant than our naive
`horizon', i.e., the effective distance traveled on the observable
brane is shorter than $h_{(1,3)}$ in the same time.

On the other hand, as shown above, the scenario of patched causal
paths (without a single smooth causal geodesic) can solve the horizon
problem.  Such a scenario may be effective when there is another
``hidden sector'' brane at $u=L$ and the intersections of segments A
and B or segments B and C represent vertices of interactions of the
bulk fields with the fields confined on the brane.  Hence, since the signal
jumps from one geodesic to another through interactions,  
the bulk fields must interact sufficiently strongly with the fields on
each of the branes for this scenario to be viable.

Although naively this requirement may seem problematic in our
scenario, in reality, sufficient interactions may be possible.  For
example, suppose one considers an action of the form
\begin{equation}
S \ni \int d^5 x \sqrt{g}  \partial_\mu \phi
\partial^\mu \phi  +
\int_{\mbox{brane 1}} \lambda_1 \phi \bar{\psi} \psi +
\int_{\mbox{brane 2}} \lambda_2 \phi \bar{\psi} \psi +
\int_{\mbox{branes 1,2}} {\rm K.E.}(\psi)
\label{eq:naivecoupling}
\end{equation}
where $\phi$ is a massless bulk scalar field (with the dimensions
properly normalized) and $\psi$ is a four dimensional fermion
confined to the boundary.  After integrating out the $\phi$ field
we obtain interactions of the form
\begin{eqnarray}
S_{\mbox{eff}} & \ni & \int_{\mbox{brane 1}} d^{4}x_1 s_1
\left[\frac{\lambda_1^2 s_1}{4} \bar{\psi}(x_1) \psi(x_1)
\int_{\mbox{brane 1}} d^{4}x_1' G(x_1, x_1') \bar{\psi}(x_1')
\psi(x_1') \right. \nonumber \\ & & \left.  + \frac{\lambda_1
\lambda_2 s_2}{4} 
\bar{\psi}(x_1) \psi(x_1) \int_{\mbox{brane 2}} d^{4} x_2' G(x_1,
x_2') \bar{\psi}(x_2') \psi(x_2') \right]
\label{eq:effectiveint}
\end{eqnarray}
where $s_1= a^3(t)$ is the ratio of the $d$ dimensional volume measure to the
extra 1 dimensional volume measure evaluated on brane 1 while $s_2 =
e^{-3 k L} a^3(t)$
is the same ratio on brane 2: i.e.
\begin{equation}
s_i = \left. \frac{\sqrt{|g|}}{\sqrt{|g_{44}|}} \right|_{\mbox{brane i}}.
\end{equation}
$G$ is the Green function of a five dimensional Klein Gordon
operator: i.e.
\begin{equation}
\frac{1}{2} \left( \partial_a \left\{ \sqrt{|g|} g^{ab}\partial_b \right\} +
\epsilon^2 \right) G(x,x') = \delta^{(5)}(x -x')
\end{equation}
where the derivatives are with respect to $x$ and the infrared
regulating mass $\epsilon$ is arbitrarily small.  Because of the $s_2$
suppression factor, one would naively expect the couplings connecting
the $\psi$s on two different branes to be suppressed.  However,
because in that case the Green function $G$ contains a $1/s_2$
behavior in the $\epsilon \rightarrow 0$ limit, the $s_2$ factors
approximately cancel and the coupling is unsuppressed.\footnote{A
fuller exploration will be given in a related work \cite{prep}.}

Finally, we note that this patched geodesic model can easily be
generalized to spacetimes with dimension greater than just five.

\subsection{Example in 2+1 Dimensions}
\label{sec:embedd21}
\begin{figure}
\hspace*{60pt} \epsfxsize=125pt \epsfbox{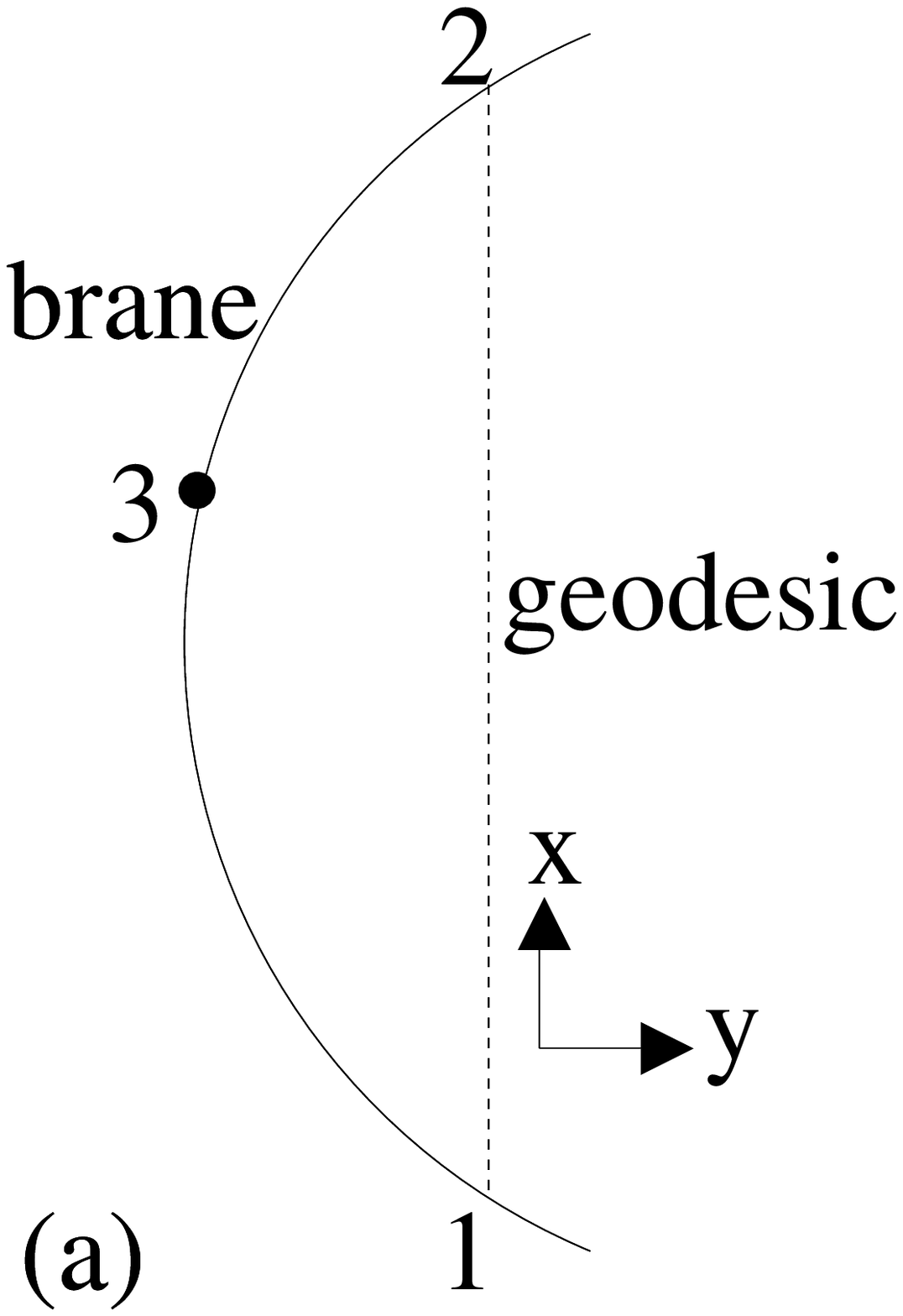}
\hspace*{120pt} \epsfxsize=125pt \epsfbox{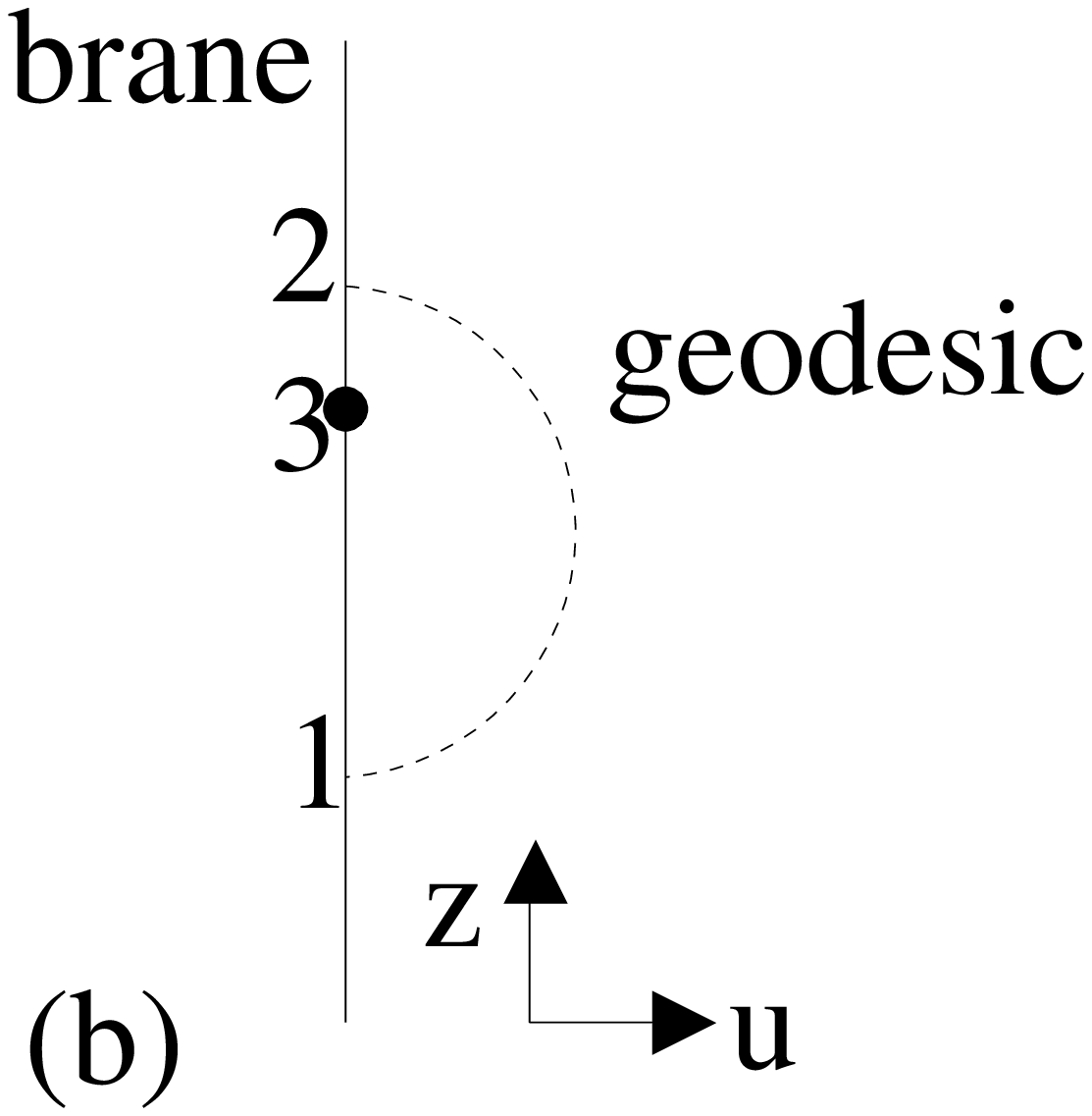}
\caption{Brane and geodesics shown in coordinate systems (x,y) and (u,z)
in 2+1 dimensional example.
The location of our brane (our observable universe) is shown as
$y=\xi(x)$ in fig. 1a and as $u=0$ in fig. 1b.  The geodesic of the full
metric is a straight line between points 1 and 2 in fig. 1a and
a curve in fig. 1b.  The distance $z_{(1,3)}$ between points 1 and 3
in fig. 1b is the horizon distance usually calculated in cosmology
in the absence of extra dimensions. Note that 
$z_{(1,2)} > z_{(1,3)}$, such that points traditionally ``outside
the horizon" are here causally connected.}
\end{figure}

For pedagogical purposes we will here describe a lower dimensional
example, in which our observable universe is a one-dimensional surface
in a spatially two-dimensional world.  We consider a Minkowski
spacetime of the form \begin{equation} \label{mink2} {\rm d s}^2 =
{\rm dt}^2 - {\rm dx}^2 - {\rm dy}^2 \, .  \end{equation} Under the
coordinate transformation \begin{equation} \label{zcoord} z= \int
\sqrt{1+[\xi'(x)]^2} {\rm dx} \, \end{equation} and \begin{equation}
\label{ucoord} u= y - \xi(x)\, , \end{equation} where the function
$\xi(x)$ will be chosen below and the prime operation (') refers to
derivatives with respect to $x$, the line element transforms to
\begin{equation} {\rm ds}^2 = {\rm dt}^2 - {\rm dz}^2 - {\rm du}^2 -
{2 \xi'(x({\rm z})) {\rm du} {\rm dz} \over \sqrt{1 + [\xi'(x({\rm
z}))]^2}} \, .  \end{equation} We will choose the location of the
brane (that is our observable universe) to be at $u=0$, such that in the
(u,z) coordinate system, the brane is merely a straight line with $z$
as the coordinate on the brane.  From eqn.(\ref{ucoord}), one can see
that, in the $(x,y)$ coordinate system, the location of the brane is
at y$=\xi(x)$.  Hence, in this coordinate system, the brane looks
curved.  Figure 2a,b shows the location of the brane in the $(x,y)$
and $(u,z)$ coordinate systems respectively. Since the $(x,y)$
coordinate system is trivial (Minkowski), it is obvious that a
geodesic is simply a straight line.  We have hence plotted such a
geodesic in Figure 2a between two points 1 and 2.  This same geodesic
(of the full metric) becomes a curved line in the $(u,z)$ coordinate
system, as shown in Figure 2b.  For comparison, in figure 1b we have
also plotted a third point 3, which is a geodesic of the induced
metric on the brane.  It is the distance $z_{(1,3)}$ that is the usual
horizon that we calculate in cosmology when geodesics off the brane
are not considered.  The claim is that, in the same amount of time,
the distance traveled by a null signal (e.g., light ray) directly
along the brane from point 1 to point 3 is shorter than the effective
distance on the brane traveled by a light ray traveling from point 1
to point 2.  Hence, while we might naively think points 1 and 2 are
causally disconnected, in fact they are causally connected by
information traveling off the brane at point 1 and reentering the
brane at point 2.  This behavior can be explained because, in the
trivial Minkowski metric of the $(x,y)$ coordinates, the distance
between points 1 and 2 can be traversed by a straight line while the
distance between points 1 and 3 must involve a curved and hence longer
path.

We will illustrate the above quantitatively.  In the $(u,z)$ coordinate system,
a particle moving along the brane from points 1 to 3 satisfies
\begin{equation}
0 = {\rm ds}^2 = {\rm dt}^2 - {\rm dz}^2
\end{equation}
such that the distance $z_{(1,3)}$ traveled along the brane 
between points 1 and 3 in time $t_f$ is
\begin{equation}
z_{(1,3)} = t_f \, .
\end{equation}
This is the distance that one would naively calculate as the horizon
size in eqn.(\ref{sizeobs}).  However, let us consider a 
geodesic that leaves and subsequently reenters the brane.
For simplicity, we can choose $y$ to be a constant $y_1$
for the particular geodesic we consider (as drawn in Figure 2a).

Then, in the (y,x) coordinate system, the geodesic is given by
\begin{equation}
\label{geod1xy}
y_g =y_1,   \,\,\,\, x_g= - c + t \, , 
\end{equation}
where $c$ is a positive constant, subscript $g$ refers
to the geodesic, and we assume the signal leaves
the brane at initial time $t=0$.  

Now to proceed we will choose a particular form for $\xi(x)$,
\begin{equation}
\xi'(x) = {\rm tan(kx)} \, ,
\end{equation}
i.e.,
\begin{equation}
\xi(x) = -{1 \over k} {\rm ln(cos(kx))} \, ,
\end{equation}
where $k$ is a constant.
Then, using eqns.(\ref{zcoord}) and (\ref{ucoord}), 
we can transform the geodesic in eqn.(\ref{geod1xy}) to the
(u,z) coordinate system:
\begin{equation}
u_g(t) = y_1 - \xi(t-c) = y + {1 \over k} {\rm ln}[{\rm cos}k(t-c)]\, ,
\end{equation}
\begin{equation}
\label{geod1z}
z_g(t) = {1 \over k}{\rm ln}[{\rm sec}k(t-c) + {\rm tan}k(t-c)] \, .
\end{equation}
Hence $z_{(1,2)} = z_g({t_f})$ is the z-coordinate
distance traversed by the light ray following a geodesic 
in the full metric from points 1 to 2 in the time $t_f$.
It is clear that the second term in eqn.(\ref{geod1z}) can blow up,
such that it is certainly possible that
\begin{equation}
z_{(1,2)} > z_{(1,3)} \, ,
\end{equation}
such that seemingly causally disconnected points 1 and 2 have in fact
been in causal contact.

In the next subsection, we will generalize the continuous metric of this
subsection to an expanding universe in higher dimensions.

\subsection{Generalization of the 2+1 example to 4+1 dimensions}
\label{sec:onecurve}
Here we consider a 4+1 dimensional generalization of the 2+1
dimensional example and include cosmological expansion.  However, as
we will see, because the origin becomes a special (even singular)
point, this example is not homogeneous and hence does not describe the
real universe.  Still, it solves the horizon problem in a novel way
and as such is instructive.

Consider a metric of the form
\begin{equation}
ds^2 = dt^2 - a^2(t) ( dr^2 + f(r) d \Omega^2 + du^2 + 2 g(r) du dr)
\label{eq:fullmetric}
\end{equation}
where
\begin{eqnarray}
f(r) & = & \frac{1}{k^2} \left( \arccos \left[ \mbox{sech}(kr) \right]
\right)^2 \\ 
g(r) & = & \tanh(kr)
\end{eqnarray}
where $k$ is an arbitrary constant and $d\Omega^2 = \sin^2 \theta
d\phi^2 + d\theta^2$ is the angular metric of a 2-sphere.  Consider
the spatial dimensions of the usual observable universe (3-brane) to
be at $u=0$.  The induced metric on the 3-brane is
\begin{equation}
ds^2 = dt^2 - a^2(t) ( dr^2 + f(r) d \Omega^2)
\label{eq:induced}
\end{equation}
which is isotropic only about one particular point in general
(i.e. inhomogeneous).  Classically, the causal region is bounded by
the geodesics of a massless particle satisfying the geodesic equation
given in \eqr{eq:geodesic}.

Along the brane, the induced metric \eqr{eq:induced} implies a
geodesic that is different from the geodesic implied by the higher
dimensional embedding metric \eqr{eq:fullmetric}.  For the embedding
metric, the geodesics leave the brane initially and reintersect the
brane at a later time.  Suppose we consider
a light signal starting from the origin at the big bang singularity
(when a(t=0)=0).  The geodesic on the brane is then
\begin{equation}
r_b(t)=\int_0^{t} \frac{dt}{a}
\end{equation}
and the comoving horizon length (for example at the time of
nucleosynthesis) is $ L_n = r_b(t_n)$.  In the embedding higher
dimensional spacetime, the geodesic is given by
\begin{eqnarray}
r_g(t)& = &\frac{1}{k} \ln \left\{ \sec \left[ k
\frac{r_b(t)}{\sqrt{1+c^2}} \right]+
\tan \left[ k \frac{r_b(t)}{\sqrt{1+c^2}} \right] \right\} \\
u_g(t) & = & c r(t) + \frac{1}{k} \ln \left\{ \cos[k
\frac{r_b(t)}{\sqrt{1+c^2}}] \right\} \\
c & = & \frac{ \ln\{ \cosh[kr_b(t_f)]\} }{\arccos\{\mbox{sech}[k r_b(t_f)]\}}
\end{eqnarray}
where $t_f$ is the time at which the geodesic reintersects the brane
after leaving the brane at the initial time $t=0$.  If we set
$t_f=t_n$ and vary $k$ appropriately, we can make $L_n=r(t_n)$ of
\eqr{eq:horizonlength}  arbitrarily
large.  Hence, points which from the brane point of view are outside of
the causal horizon are actually causally connected.

The metric given by \eqr{eq:fullmetric} is not realistic because the
cosmology on the brane is not homogeneous.  Indeed, the result is not
surprising since the spatial curvature on the noncompact brane is
positive.  The only homogeneous constant curvature 4-dimensional
manifolds are the 3 types of FRW metric (positive, zero, and negative
intrinsic spatial curvature).  The only boundariless positive
curvature object of constant curvature in 3+1 dimensions is a
3-sphere, which is compact.  Hence, we were doomed to begin with in
trying to construct a noncompact homogeneous cosmology by embedding a
curved manifold in an Euclidean space.  The reason why this approach was
successful for the 2+1 embedding of a 1+1 dimensional manifold was the
fact that the intrinsic spatial curvature is always 0 for a one
dimensional manifold.

\section{Physical Scenario}
\label{sec:physical}
In this section we discuss a number of issues regarding the
apparent causality violation attending the scenarios we have discussed.

Firstly, let us consider whether bulk fields in such higher
dimensional spacetimes will contradict any observations.  In order for
the geodesic in the higher dimensional spacetime to solve the horizon
problem, causality must be apparently violated within the 3-brane, at
least on cosmological scales during some early time.  In order for the
apparent causality violation to be hidden today while still solving
the horizon problem, the geodesic through the extra dimension must not
be accessible today.  This is possible, for example, if the form of
the cosmological energy density early on in the universe had bulk
field coupling, while the form of the cosmological energy density
today in the universe has no bulk field coupling.

Secondly, we remark that the apparent causality violation during
an early epoch of the universe can be understood in terms of
a higher dimensional Green function that falls off less rapidly
than one would naively expect in the absence of extra dimensions.
The causality violation can manifest in terms of nonlocal interactions
of the effective 4-dimensional Lagrangian.  As in the last section,
consider for example the interaction given by
\begin{equation}
S \ni\int_{\mbox{brane}} \lambda_1 \phi \bar{\psi} \psi.
\end{equation}
As before, after integrating out the bulk field $\phi$ we find the
effective interaction given by
\begin{equation}
S_{\mbox{eff}}  \ni  \int_{\mbox{brane}} d^{4}x_1 s_1
\frac{\lambda_1^2 s_1}{4} \bar{\psi}(x_1) \psi(x_1)
\int_{\mbox{brane}} d^{4}x_1' G(x_1, x_1') \bar{\psi}(x_1')
\psi(x_1')  \, .
\end{equation}  
For $x_1'$ that is outside of the 4-dimensional light-cone of $x_1$,
we would generally expect the Green function $G$ to fall off
exponentially.  However, in the full 5-dimensional spacetime, the
point $x_1'$ that is outside of the 4-dimensional light-cone of $x_1$
will still be within the 5-dimensional light-cone, and hence the
interaction will not be exponentially suppressed.  Thus at a point
$x_1$ on the brane, one can get a contribution from another point
$x_1'$ that is farther away than usually allowed by causality, because
the 5-dimensional Green function doesn't fall off as fast outside the
4-dimensional light cone as one would naively expect in the absence
of the extra dimensions.  One can think of this as a propagator that
can leave the brane and hence connect two distant points on the brane.

Thirdly, let us consider what kind of stress energy tensor gives rise
to the metric presented in Section \ref{sec:patches}.  Note that in
light of Ref.\ \cite{chungfreese}, because the $tt$ component and the
$uu$ component of the metric are identical and time independent, we
have a fine tuned solution.  More explicitly, to support the brane
solutions, the five dimensional metric must satisfy the Israel
condition boundary conditions (see for example Refs.\
\cite{chungfreese,binetruy,chamblin} and references therein):
\begin{equation} -6 \partial_u \alpha = \kappa_5^2 \rho \\ \partial_u
\nu = \kappa_5^2( \frac{1}{2} P +\frac{1}{3} \rho) \end{equation} for
the metric of the form $ds^2 = e^{2 \nu} d\tau^2 - e^{2 \alpha} d{\bf
h}^2 - du^2$ where $\alpha = - k u + \log a(t)$ and $\nu=0$ for the
metric of \eqr{eq:simplesol}.  The pressure $P$ and the energy density
$\rho$ are associated with the fields confined on the brane.  As it
stands, the energy density of the fields confined to the brane is a
constant $\rho = -\frac{3}{2} P = 6 k/\kappa_5^2$.  However, as is
done in \cite{terning}, one can add perturbations to the brane energy
density $\rho$ such that one can obtain a component of the energy
density that dilutes as the universe expands.  Note that since we must
have $kL \approx {\cal O}(10)$ to solve the horizon problem, we must
have a brane energy density of $\rho \sim {\cal O}(100) M_5^3/L$ where
$M_5$ is the five dimensional Planck's constant defined by
$\kappa_5^2=1/M_5^3$.  Since we require $\rho < M_5^4$, this scenario
is viable only if $L \gg 100/M_5$ which is not unrealistic.  We leave
a more complete study of viable cosmologies to a future study.

For completeness, we list the rest of the Einstein equation specifying
the bulk stress energy tensor for this scenario.
\begin{eqnarray}
T^0_0 & = &- 6 k^2 + 3 (\frac{\dot{a}}{a})^2 \nonumber \\
T^1_1 & = & -3 k^2 + (\frac{\dot{a}}{a})^2 + 2 \frac{\ddot{a}}{a}
\nonumber  \\
T^4_4 & = & -3 k^2 + 3 (\frac{\dot{a}}{a})^2 + 3 \frac{\ddot{a}}{a}
\nonumber \\
T^4_0 & = & - 3 k \frac{\dot{a}}{a}
\end{eqnarray}
where we normalize the stress energy as ${\cal R}_{MN} - 1/2 g_{MN}
{\cal R } = T_{MN}$.  There are no fundamental constraints on the bulk
stress energy tensor such that these equations cannot be satisfied.

\section{Conclusion}
\label{sec:conclude}

In this paper, we have demonstrated that, in theories with extra
dimensions in which our ``observable'' 4-dimensional universe is
confined to a submanifold, there may generically exist a
non-inflationary solution to the horizon problem.  The horizon problem
can be stated as follows.  If only a finite amount of time has passed
subsequent to an initial epoch of our universe (e.g. a singularity),
then any causal signal can travel within that time period only a
finite distance, referred to as the horizon distance.  By contrast, in
the context of an FRW universe composed of ordinary matter and
radiation, there is experimental evidence that patches of the universe
which are separated by a distance longer than the horizon distance
seem causally connected.

With the existence of extra dimensions, however, the naive horizon
distance calculated by a null geodesic on the 4-dimensional
submanifold does not constitute the maximum distance a signal
can travel for a given time.  A causal signal through the extra
dimensions can reach a point in our universe which is many times
further away than the naive horizon distance.  An example of such a
higher dimensional universe is described by \eqr{eq:simplesol} with 
two 3-branes, where one of the 3-branes is our observable universe and
the other 3-brane is a ``hidden'' universe.  The field confined to our
brane can interact with the field living on a second ``hidden'' brane
a distance $L$ away from us in the extra dimension via a bulk
field.  For a given time, a causal signal can travel much further on
the ``hidden'' brane in a direction parallel to brane.  Hence, an
impulse originating on our brane can take a shortcut through the
``hidden'' brane and affect our brane at a point outside of the naive
``horizon''.  

Once ``equilibration'' of the energy density fluctuations is
established, the fields confined on the brane may decay to fields that
interact less strongly with the bulk fields.  Hence, any apparent
causality violation occurring through the existence of an extended
higher dimensional light-cone may be hidden today.

We have also studied the construction of a metric that solves the
horizon problem by embedding a curved 3-brane inside a Minkowskian
5-dimensional spacetime.  We find that, for a non-compact 3-brane, the
universe thus obtained is inhomogeneous.

\acknowledgements{DJHC thanks
Ren-Jie Zhang and Lisa Everett for useful conversations.  We thank the
Department of Energy for funding this research at the University of
Michigan. K. Freese also thanks CERN in Geneva as well as the Max
Planck Institut fuer Physik in Munich for hospitality during her stay.
} 


\end{document}